\documentclass[12pt]{article}
\usepackage{epsfig}
\usepackage{float}

\usepackage{hyperref} \hypersetup{ colorlinks=true, linkcolor=blue, filecolor=magenta, urlcolor=cyan, }
 
\def\beq{\begin{equation}}
\def\eeq#1{\label{#1}\end{equation}}
\def\eeqn{\end{equation}}

\def\beqa{\begin{eqnarray}}
\def\eeqa#1{\label{#1}\end{eqnarray}}
\def\eeqan{\end{eqnarray}}

\def\overbar#1{\overline{#1}}

\let\bar=\overbar

\def\Dslash{\not{\hbox{\kern-4pt $D$}}}
\def\dslash{\not{\hbox{\kern-2pt $\del$}}}

\def\msb{{\bar{\ssstyle M \kern -1pt S}}}

\def\Title#1{\begin{center} {\Large {\bf #1} } \end{center}}

\begin{document}

\Title{Current Status for the Inclusive Neutral Current $\pi^{0}$ production Cross Section Measurement with the NOvA Near Detector}

\bigskip\bigskip


\begin{raggedright}  

{\it Daisy Kalra\index{Kalra, D.}\\
on Behalf of the NOvA Collaboration\\
Department of Physics, Panjab University\\
Chandigarh-160014, INDIA\\
Fermi National Accelerator Laboratory, USA\\
}

\center{Talk presented at the APS Division of Particles and Fields Meeting (DPF 2017), July 31-August 4, 2017, Fermilab. C170731}

\bigskip\bigskip
\end{raggedright}

\section{Introduction}
The neutrino-Nucleus ($\nu$-N) interactions have been studied intensively for decades~\cite{Deborah}. $\nu_{\mu}$ induced neutral current (NC) interactions with a $\pi^{0}$ in the final state are the dominant background for experiments looking for the $\nu_{e}$ appearance such as NOvA and DUNE~\cite{Kevin}. The signal for the $\nu_{e}$ appearance channel is an electron in the final state that showers electromagnetically. Neutral pions decay into two photons can fake the $\nu_{e}$ appearance signal in two ways: either 2 $\gamma$'s can merge together or one of them may escape detection and hence behave like an electron shower. Therefore, a complete understanding of NC $\pi^{0}$ production is very important.

\begin{figure}[htb]
\begin{center}
\epsfig{file=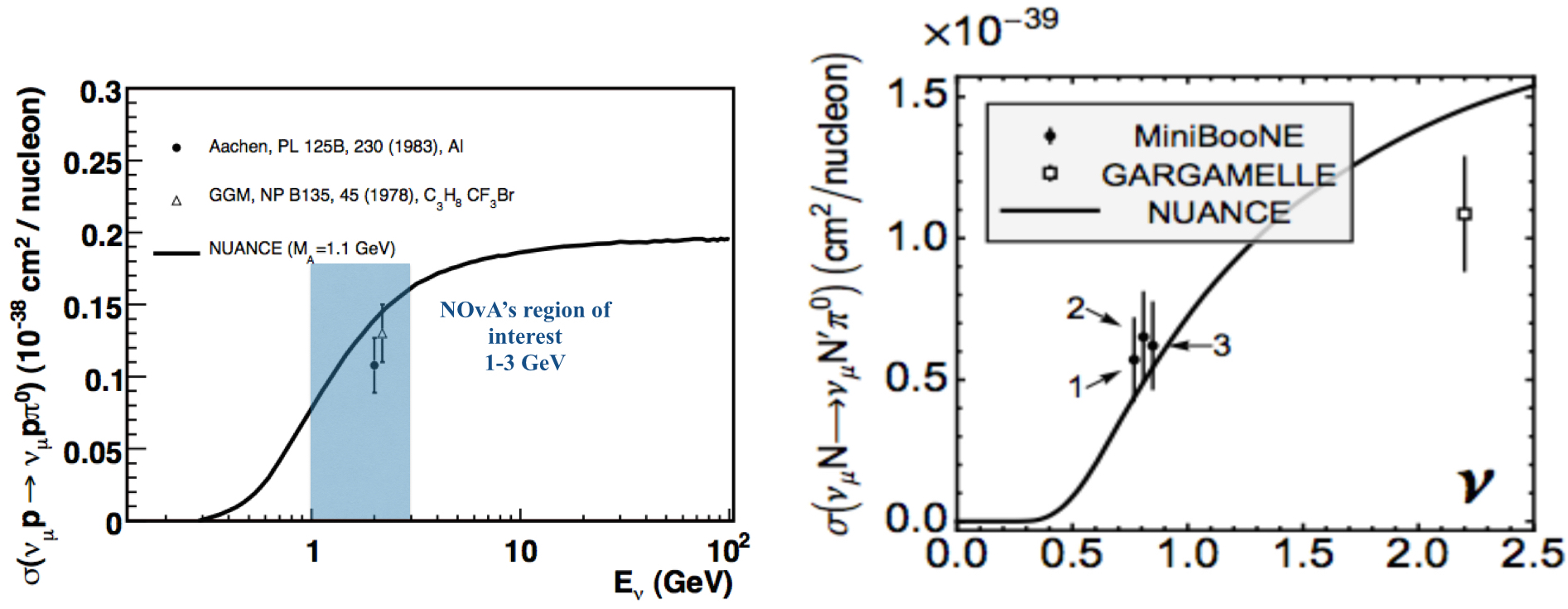,height=2.0in, width=0.8\textwidth}
\caption{Left: Existing measurement of the cross section for the NC process, $\nu_{\mu}$p$\rightarrow$ $\nu_{\mu}$p$\pi^{0}$, as a function of neutrino energy~\cite{Formaggio}. Right: The flux-averaged total cross sections for $\nu_{\mu}$ induced incoherent exclusive NC 1$\pi^{0}$ production on $CH_{2}$ in MiniBooNE corrected for FSI. Points 1, 2 and 3, are the cross sections extracted using different models~\cite{Arevalo}. }
\label{fig:ncpi0data}
\end{center}
\end{figure}

At present, only a few measurements of neutrino-induced $\pi^{0}$ production have been performed (Figure ~\ref{fig:ncpi0data} (left)). Most of this data is in the form of ratios (neutral current (NC) / charged current (CC))~\cite{Formaggio}. There also exists the absolute cross section measurement around 1 GeV for incoherent $\pi^{0}$ production and an inclusive cross section measurement from the MiniBooNE experiment(Figure~\ref{fig:ncpi0data} (right))~\cite{Arevalo}. In this paper, we present the current status for the inclusive neutral current  $\pi^{0}$ production cross section with the NOvA Near Detector (ND).

\section{The NOvA experiment}
NOvA stands for NuMI off-axis $\nu_{e}$ appearance. It is a long-baseline accelerator-based neutrino oscillation experiment designed to study primarily the $\nu_{e}$ appearance using neutrinos ($\nu_{\mu}$) from the Main Injector (NuMI) beam ~\cite{Adamson}. In addition to measuring the $\nu_{e}$ appearance rate, we also measure the $\nu_{\mu}$ disappearance rate (via $\nu_{\mu} \rightarrow \nu_{\mu}$ disappearance channel) with both neutrinos and anti-neutrinos. The NOvA experiment, studying these channels, aims to resolve the neutrino mass hierarchy, determine the CP-violating phase ($\delta_{CP}$) and $\theta_{13}$, $\theta_{23}$ octant (muon-tau asymmetry in neutrino mixing) and precisely measure the atmospheric parameters ($\theta_{23}$ and $\delta m^{2}_{atm}$) using two detectors~\cite{AdamsonI}~\cite{AdamsonII}.

\subsection{The NOvA detectors}
NOvA studies $\nu$-oscillations using two detectors- The NOvA near detector (ND) and The NOvA far detector (FD). The 193 ton ND has dimensions 3.9 m $\times$ 3.9 m $\times$ 12.67 m and is located 100 m underground at Fermilab. The 14 kton FD has dimensions 15.5 m $\times$ 15.5 m $\times$ 60 m and is located approximately 810 km away on the surface in Ash River, Minnesota. Both the detectors are sited 14.8 milliradians off the NuMI beam axis. The off-axis position helps in reducing the NC background while maintaining a high neutrino ($\nu_{\mu}$) flux peaked around 2 GeV in energy, where it gives the highest probability of oscillations. Both the detectors use identical technology which helps to reduce beam/flux related systematics in the oscillation measurements.

\begin{figure}[htb]
\begin{center}
\epsfig{file=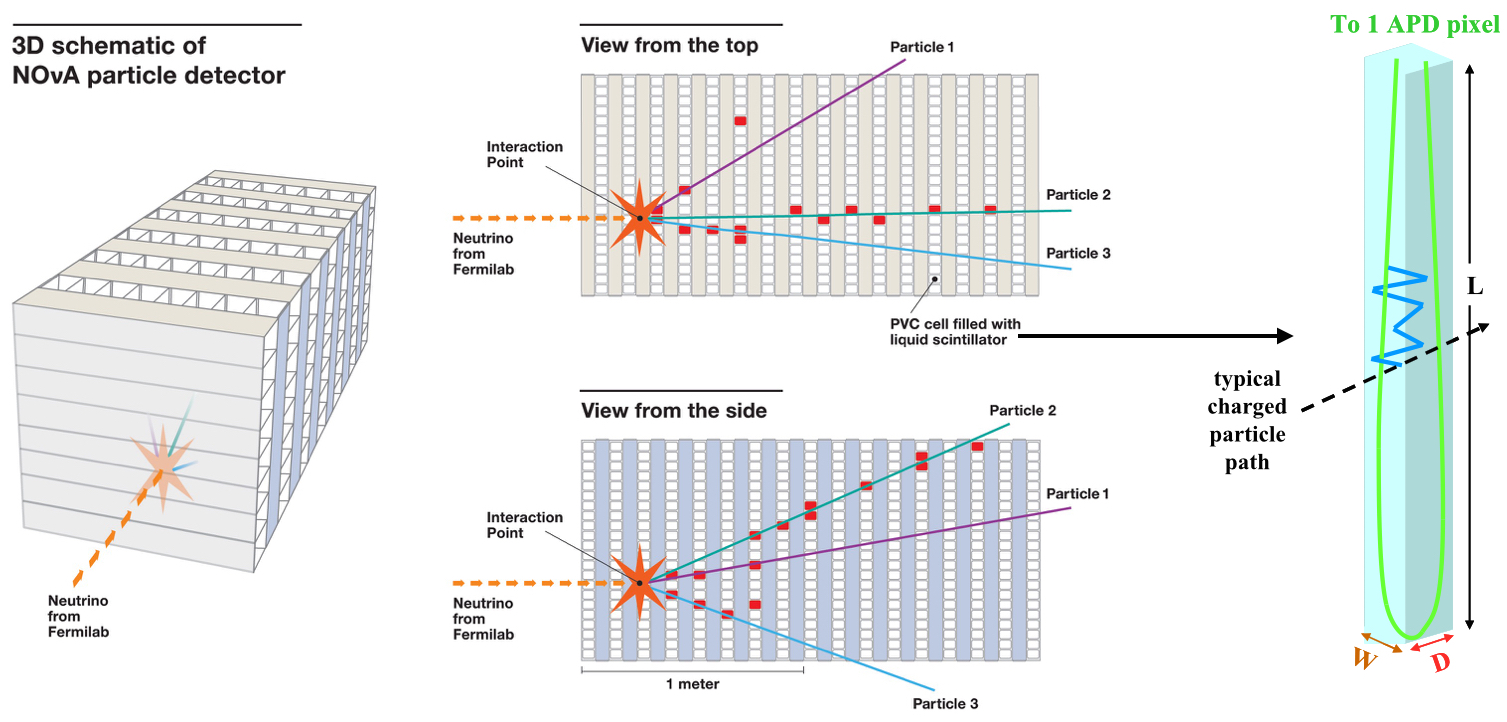,height=2.0in, width=0.8\textwidth}
\caption{Leftmost portion shows a 3D view of the detector. Middle portion shows a neutrino interaction in both views of the detector. Rightmost portion shows a PVC plastic cell containing liquid scintillator and WLS fiber to collect scintillating light and routes the light to an avalanche photodiode (APD).}
\label{fig:detector3dview}
\end{center}
\end{figure}

NOvA uses polyvinyl chloride (PVC) plastic cells extrusions~\cite{Talaga} filled with liquid scintillator and wavelength shifting (WLS) fibers. The extrusions are 15.5 m long in the FD and 3.9 m long in the ND. Each cell is 3.9 cm wide and 6.6 cm deep (in cross section).  Cells are arranged into 32-cell modules which are stacked to form a plane. There are alternate planes with horizontally and vertically oriented cells that allows for a 3D reconstruction. Vertical planes form the "top view" (XZ view) of the detector and horizontal planes for the "side view" (YZ) view of the detector. Figure~\ref{fig:detector3dview} shows a neutrino interaction in both views. NOvA's liquid scintillator is mineral oil with 4\% pseudocumene and comprises 62\% of the detector mass. It serves as both an interacting medium and a source of light collected by front end electronics. 
  
 When a charged particle traverses a cell, it ionizes liquid scintillator resulting in the production of scintillating light which is collected by the WLS fiber. The fiber is looped at the bottom and both ends of the fiber go to one pixel of an avalanche photodiode (APD) that converts photons to the electronic signals. 
 NOvA detectors are low Z, highly active tracking calorimeters which are able to diffrentiate between muons (long tracks), electrons (electromagnetic (EM) showers) and pions (leave a gap before decaying to gammas). NOvA's design allows for very good EM shower reconstruction to tag $\nu_{e}$, the primary goal of NOvA ($\nu_{e}$ appearance).

\subsection{The NuMI beam line}
\begin{figure}[htb]
\begin{center}
\epsfig{file=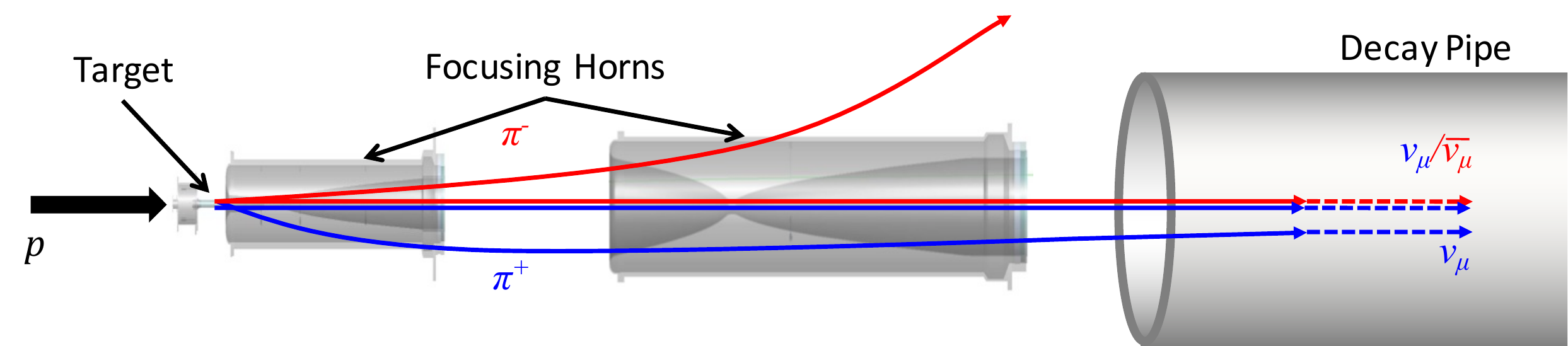,height=1.3in}
\caption{The NuMI beam line at Fermilab.}
\label{fig:numibeamline}
\end{center}
\end{figure}

The NuMI (neutrinos at Main Injector) facility at Fermilab produces a high intensity neutrino ($\nu_{\mu}$) beam. Figure~\ref{fig:numibeamline} shows a sketch of the NuMI beam line  where, 120 GeV protons from the Main Injector strike a 122.5 cm long graphite target and produce many short-lived particles such as pions, kaons. These particles are focused by a set of two magnetic horns with a horn current of +200 kA (-200 kA).  These particles then travel towards a helium filled decay pipe where they decay into neutrinos (anti-neutrinos) and muon. Depending on the horn current, we define two beam configuration: Forward Horn Current (FHC: +200 kA) and Reverse Horn Current (RHC: -200 kA). The FHC configuration focuses charged particles with positive polarity ($\pi^{+}$,$K^{+}$) which decay to give a neutrino beam ($\nu_{\mu}$) whereas, the RHC configuration focuses charged particles with opposite polarity ($\pi^{-}$,$K^{-}$) that decay to give an anti-neutrino enhanced beam ($\overbar{\nu}_{\mu}$). The FHC NuMI beam is composed mostly of $\nu_{\mu}$ with a 3.8\% $\overbar{\nu}_{\mu}$ component and 2.1\% ($\nu_{e}$+$\overbar{\nu}_{e}$) component~\cite{Adamson}.

\begin{figure}[htb]
\begin{center}
\epsfig{file=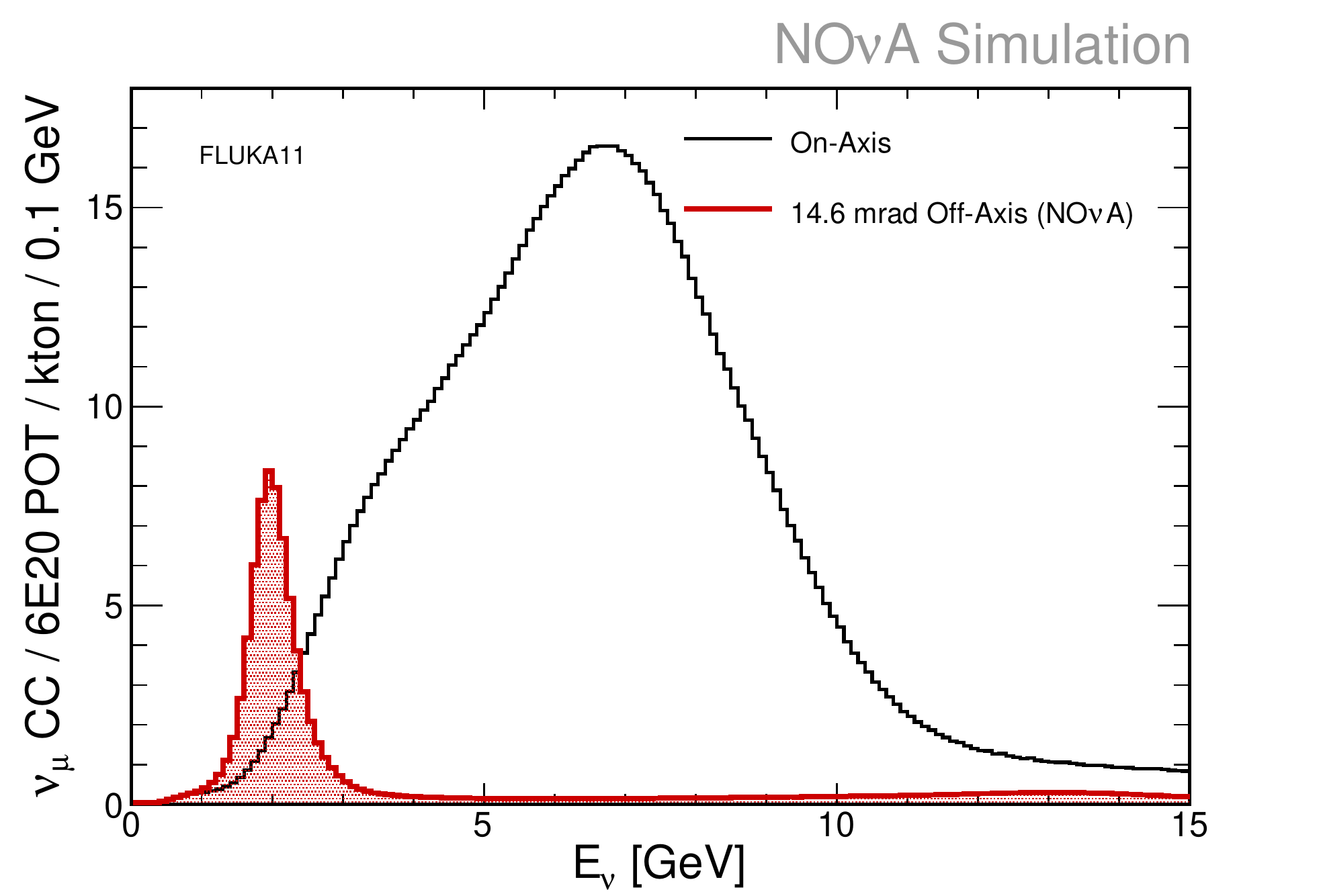,height=2.0in, width=0.5\textwidth}
\caption{Neutrino energy spectrum in the NOvA FD. The NOvA off-axis position selects a quasi mono energetic  $\nu_{\mu}$ beam of 2 GeV to have the highest probability of oscillation at the NOvA FD.}
\label{fig:fdSpectrum}
\end{center}
\end{figure}

The NOvA detectors, being sited at an off-axis position, see a high intensity $\nu_{\mu}$ beam peaked at 2 GeV in energy (red spectrum) as shown in Figure~\ref{fig:fdSpectrum}. 

\section{Monte Carlo Study}

We performed the studies presented here using ND Monte Carlo (MC) datasets. As the NOvA ND is located close to the target source (approximate 1 km away from target source), it has high statistics and thus provides an excellent opportunity for the measurement of various neutrino interactions mainly cross-section measurements. The ND MC datasets used for the following analysis have ~x4 higher statistics than the existing ND data. Neutrino interactions in the NOvA detectors are simulated using the GENIE event generator~\cite{Andreopoulos}. ND data used here were collected in FHC beam configuration. The distributions here, in this paper, reflect the data available at the NOvA ND, corresponding to $~8.09\cdot10^{20}$ Protons On Target (POT).

\subsection{Signal and Background}

We define our signal as $\nu_{\mu}$ induced NC interactions containing at least one $\pi^{0}$ in the final state with kinetic energy $>$ 0.5 GeV. The challenge is to reconstruct and identify the final state of an interaction to be able to identify the $\pi^{0}$'s decay into electromagnetic (EM) showers.

\begin{figure}[htb]
\begin{center}
\epsfig{file=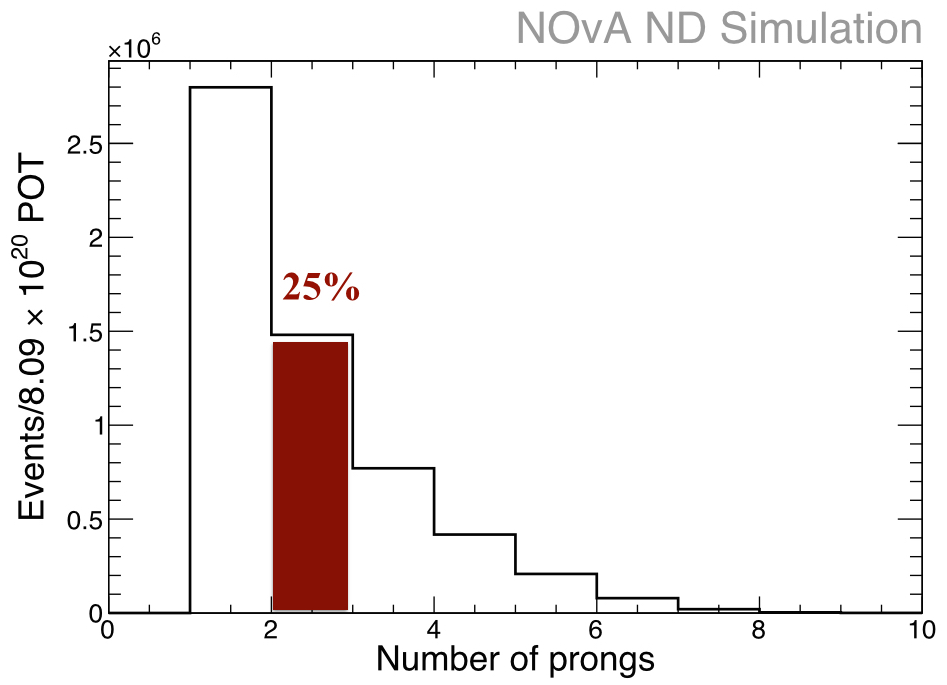,height=2.0in, width=0.5\textwidth}
\caption{Number of reconstructed prongs in the signal slices.}
\label{fig:prongs3d}
\end{center}
\end{figure}

Neutrino interactions in the NOvA ND are reconstructed into slices (clusters of cell hits that are closely related in space and time) and then these clusters are examined to find the particle paths using Hough transformation~\cite{Fernandes}. The information from the intersection of the paths is used to find a neutrino interaction vertex (a point where the primary neutrino interaction takes place). Clusters of hits that correspond to the same shower are reconstructed as prongs. The leading prong (prong1) is the most energetic prong and the sub leading prong (prong2) is the second-most energetic prong and so on. 
A first preselection is made requiring the reconstructed vertex to be in the ND fiducial volume (-180 cm$<$vertex X and Y$<$180 cm and 50.0 cm$<$vertex Z$<$1000.0 cm) and all the reconstructed showers to be contained (-180 cm$<$shower stop X,Y$<$180 cm and 200 cm$<$shower stop Z$<$1200 cm). The distribution of the number of prongs reconstructed for the signal slices is shown in Figure~\ref{fig:prongs3d}.

\begin{figure}[htb]
\begin{center}
\epsfig{file=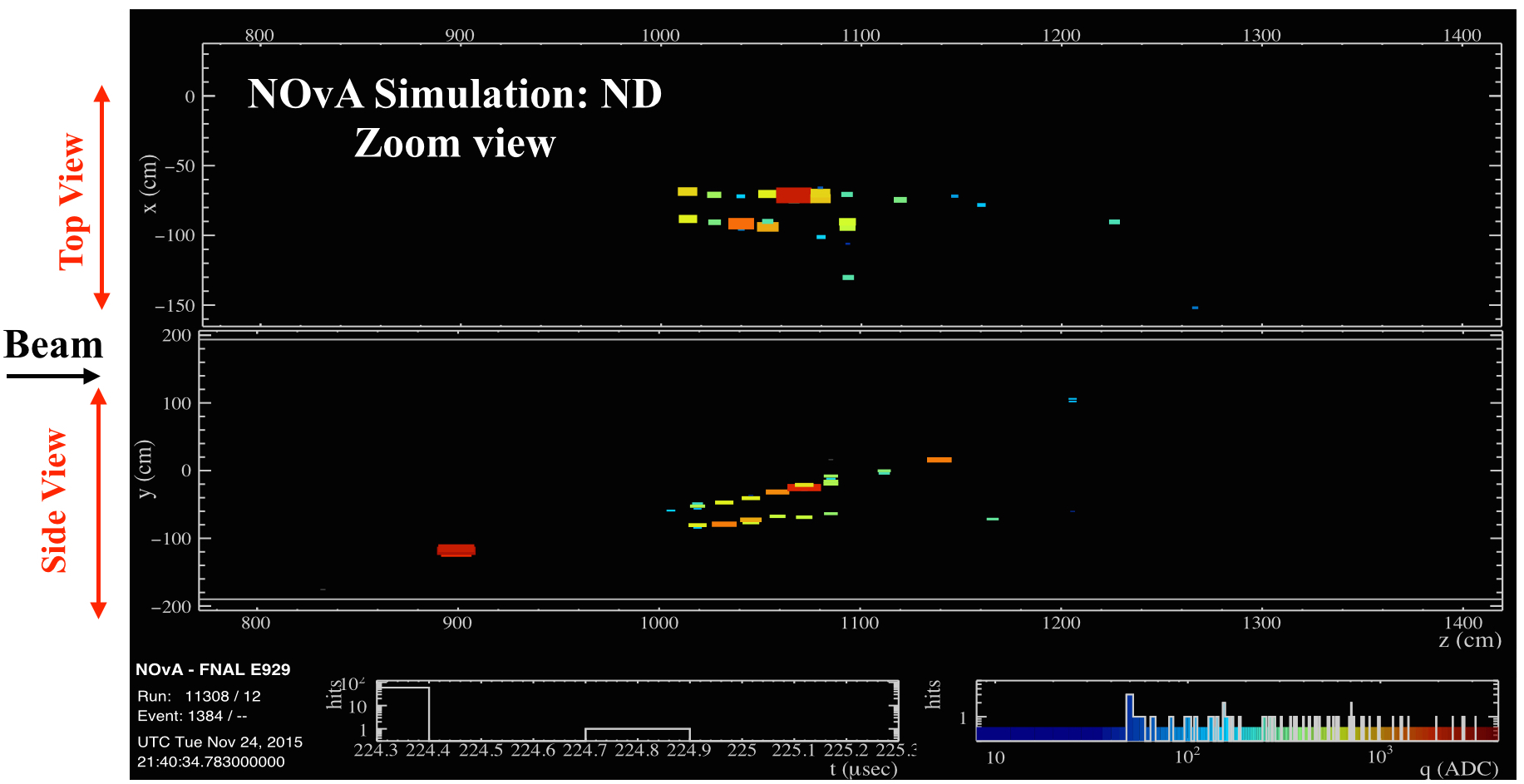,height=2.3in, width=0.8\textwidth}
\caption{Event display of the signal event with 2 reconstructed prongs.}
\label{fig:evd}
\end{center}
\end{figure}

 This analysis considers only 2-prong events (25\% of the total signal slices) originating from a common vertex. The event display in Figure~\ref{fig:evd} shows the signal event (2-prong) as simulated in the NOvA ND.
 
The background to the signal events comes from $\nu_{\mu}$ CC interactions, where the outgoing $\mu$ is not identified, with or without a $\pi^{0}$ (CC background) and  $\nu_{\mu}$ NC interactions without a $\pi^{0}$ or with a $\pi^{0}$ below kinetic energy threshold, 0.1 GeV(NC background). Intrinsic beam contamination from $\nu_{e}$ is also a source of background but is negligible. 

\subsection{Background Rejection}

This analysis has a huge amount of background events as compare to the signal events. To reject the background events, we define a series of pre-selection cuts (cuts applied prior to the final event selector). A first pre-selection is made requiring all the events vertices to be inside the ND fiducial volume, which helps in rejecting "Rock events" (the events entering from the edges of the detector). 

\begin{figure}[htb]
\begin{center}
\epsfig{file=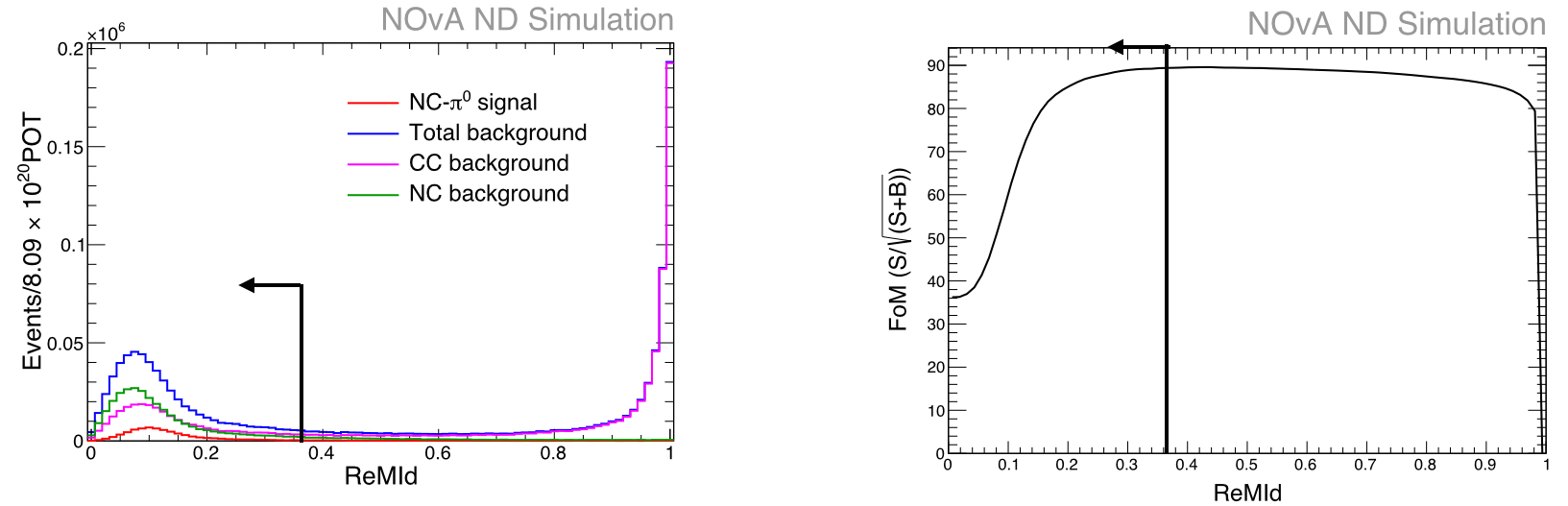,height=2.0in, width=0.9\textwidth}
\caption{Left: ReMId distribution for the signal and background events with fiducial and containment cuts. Right: The FOM distribution evaluated from the ReMId distribution.}
\label{fig:remidandfom}
\end{center}
\end{figure}

Then we include NOvA's Reconstructed Muon Identification (ReMId) variable in the pre-selection. ReMId is a particle identification (PID) algorithm specifically based on the muon tracks, that has been developed to be used in the NOvA's  $\nu_{\mu}$ disappearance analysis~\cite{AdamsonI}. It selects muons from the $\nu_{\mu}$ CC interactions and gives a value 0 to 1 to an event, where 1 is for CC-like events. Figure~\ref{fig:remidandfom} (left) is the ReMID distribution for the 2-prong events with fiducial and containment cuts that shows a huge amount of CC events, background events in this analysis, around 1. So, to reject these background events, we choose a cut value on the ReMId based on its figure of merit (FOM = S$/\sqrt(S+B)$, where S is signal events and B is background events) as shown in Figure~\ref{fig:remidandfom} (right). The FOM is maximized at 0.36, so we include ReMId $<$ 0.36 in the pre-selection.

\begin{figure}[htb]
\begin{center}
\epsfig{file=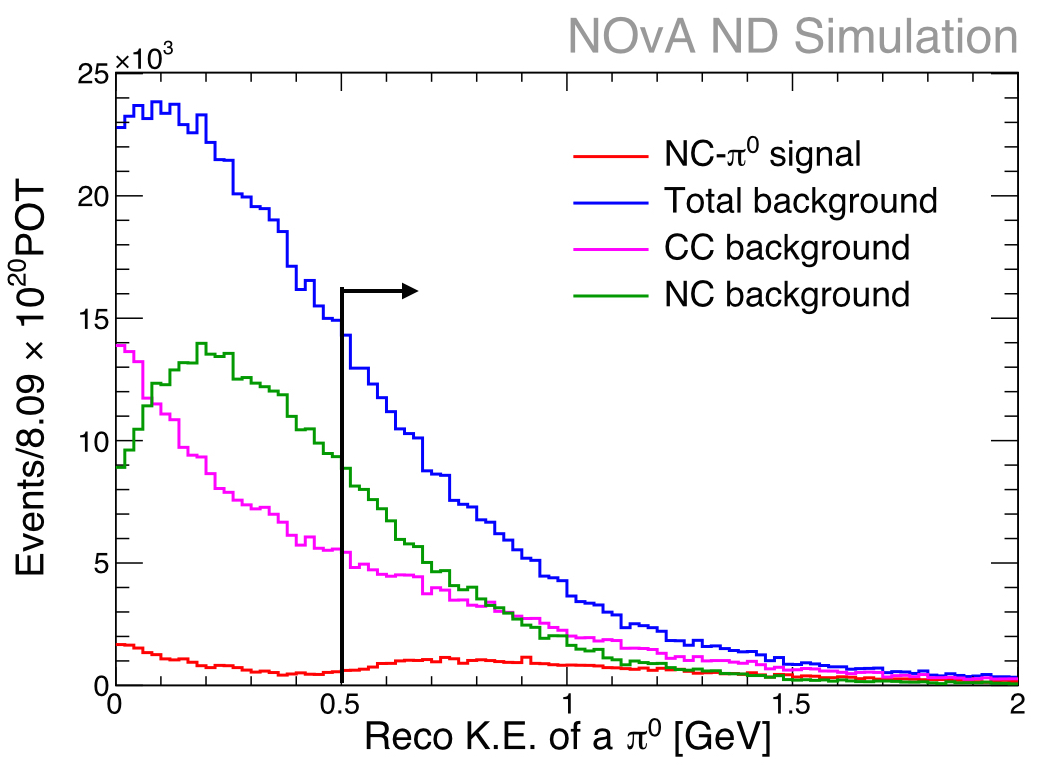,height=2.0in, width=0.5\textwidth}
\caption{Reconstructed kinetic energy of a $\pi^{0}$ variable distribution for the signal and background events with fiducial, containment and ReMId $<$ 0.36.}
\label{fig:recokinen}
\end{center}
\end{figure}

Then, we include the reconstructed kinetic energy (Reco K.E.) of a $\pi^{0}$ variable in the pre-selection.  Figure~\ref{fig:recokinen} shows the Reco K.E. distribution for the signal and background events with fiducial, containment and ReMId $<$ 0.36. Reco K.E $>$ 0.5 GeV rejects $~$80\% of the background w.r.t the background with fiducial and containment cuts and maximizes the FOM. So, we include Reco K.E $>$ 0.5 GeV in the pre-selection.

\subsection{Event Selection}

With the pre-selection (fiducial, containment, ReMId $<$ 0.36 and Reco K.E. $>$ 0.5 GeV), we studied mutivariate analysis (MVA) algorithms for effeciently selecting the 2-prong signal events. MVA works in two phases, training phase and testing phase~\cite{Hoecker}. In training phase an algorithm is trained using a set of input variables and in testing phase the training output is used to test the algorithm. For training and testing, we split the ND MC dataset where, 50\% of the dataset is used for training and another 50\% is used for testing the algorithm.  For this analysis, we compared several algorithms and based on the performance, Boosted Decision Tree- Gradient (BDTG) is chosen to be used as a final event selector. To train the BDTG, we tried various sets of input variables and selected the one with the best performance in terms of efficiency and background rejection.

\begin{figure}[htb]
\begin{center}
\epsfig{file=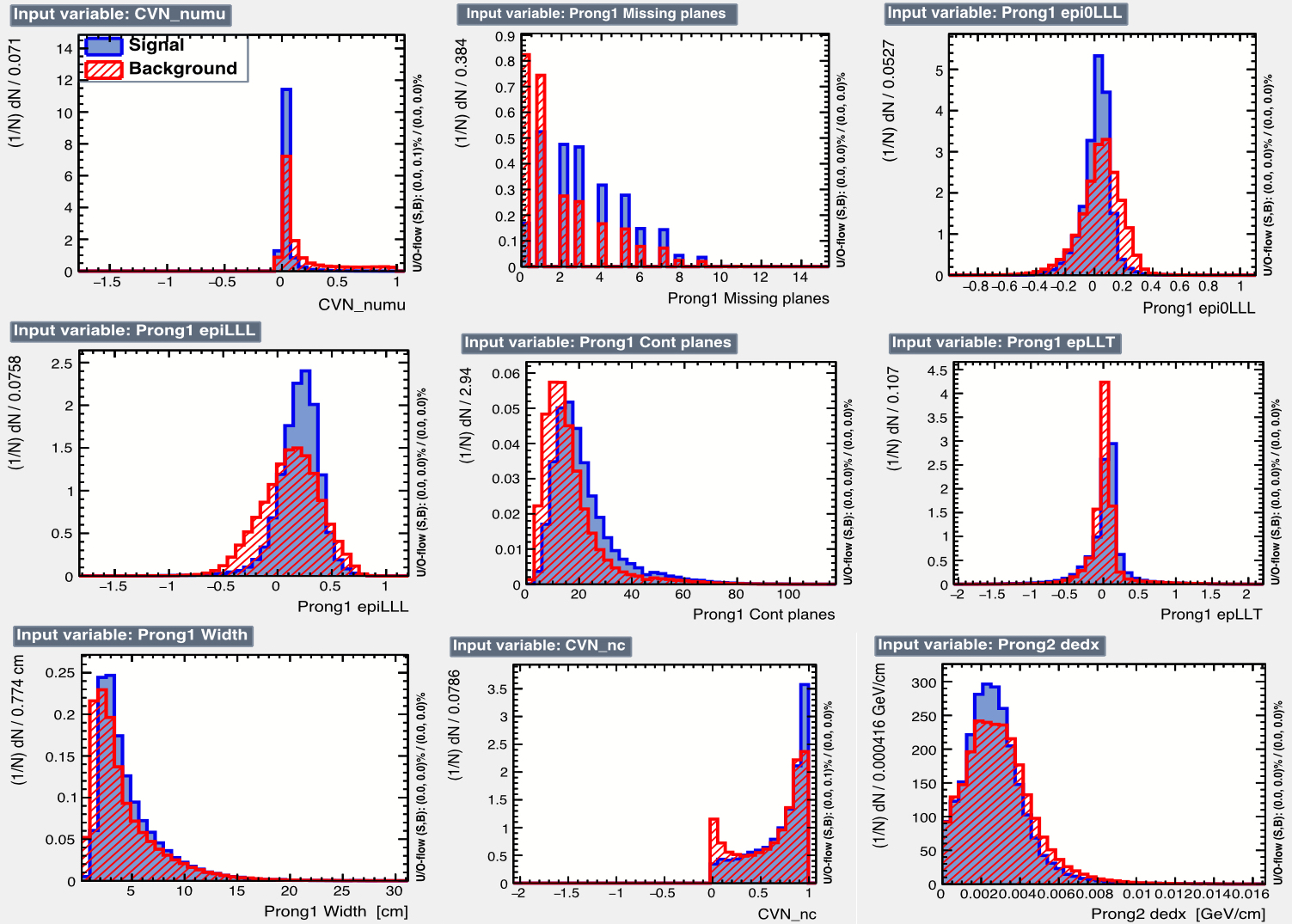,height=2.9in, width=0.9\textwidth}
\caption{Distribution of input variables for signal (blue) and background (red) events.}
\label{fig:inputvars}
\end{center}
\end{figure}

Figure~\ref{fig:inputvars} shows a distribution of input variables selected to train the BDTG. Those variables are chosen which tells us about the prong related characteristics. For instance, prong1 missing planes (number of planes without any prong1 hit), prong1 contiguous planes (number of continuous planes with the most energetic prong hits), prong1 width (shower width [cm]), prong2 $\frac{dE}{dx}$ (Average energy loss by the prong2), prong1 $e-\pi^{0}$ LLL (electron$-\pi^{0}$  log-likelihood for the longitudinal shower (a measurement is performed plane by plane) where, difference gives a measure of likelihood the shower is an electron shower compared to a $\pi^{0}$) , prong1 $e-\pi^{\pm}$ LLL (electron$-\pi^{\pm}$ log-likelihood for the longitudinal shower) and prong1 e-p LLT (electron-proton log-likelihood for the transverse shower (a measurement is performed cell by cell)). 
 
Additionally, we also use NOvA's Convolutional Visual Network (CVN) algorithm, developed to identify neutrino interactions, based on their topology~\cite{Aurisano}. Four separate CVN classifiers are developed based on interaction types, $\nu_{\mu}$ CC, $\nu_{e}$ CC, $\nu_{\tau}$ CC and $\nu$ NC. Out of these four, we use two classifiers, CVN $\nu_{\mu}$ CC and CVN $\nu$ NC, which are powerful in rejecting the background in this analysis. Interaction type $\nu_{\mu}$ CC has a muon and hadronic component in the final state which is characterized by long, low dE/dx track whereas, in $\nu$ NC interaction the final state has a neutrino (can't detect) and visible hadronic component.  Both the classifiers gives a value 0 to 1 to an event where, 1 is for NC like events in CVN $\nu$ NC and $\nu_{\mu}$ CC like events in CVN $\nu_{\mu}$ CC.

\begin{figure}[htb]
\begin{center}
\epsfig{file=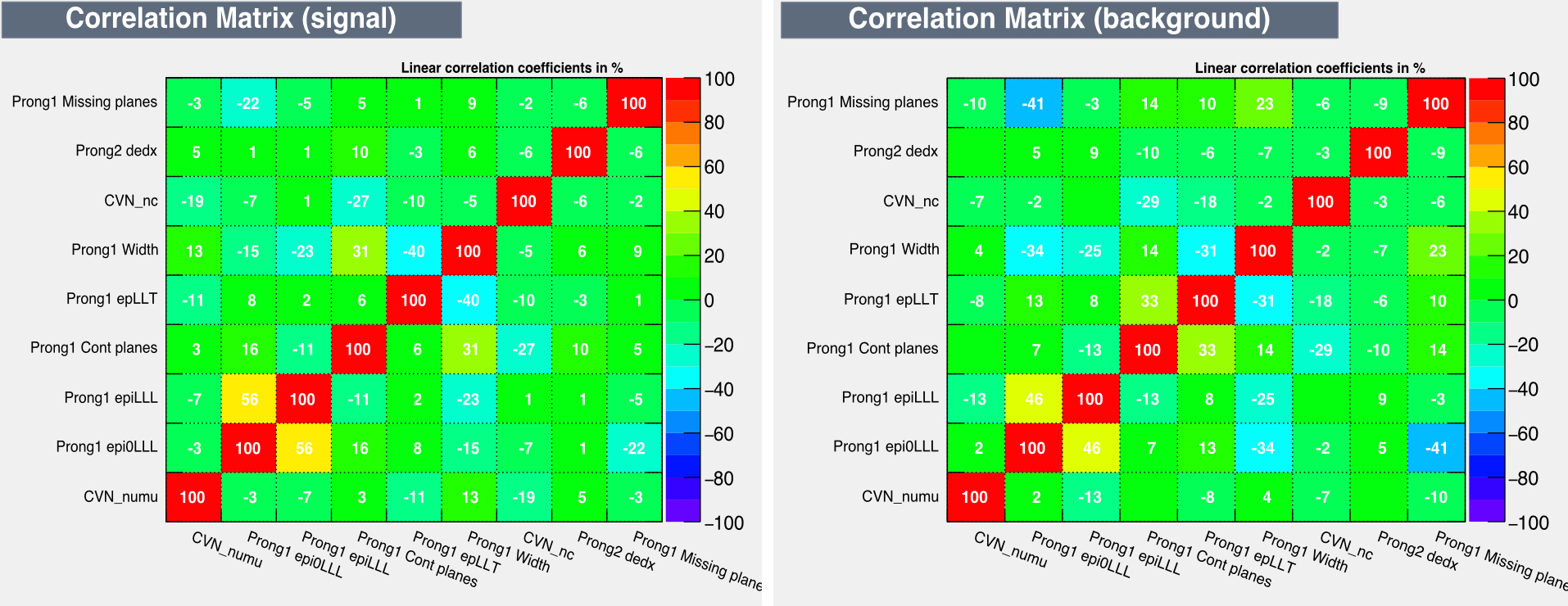,height=2.0in, width=0.8\textwidth}
\caption{Correlation matrix of input variables for the signal and background events.}
\label{fig:cormatrix}
\end{center}
\end{figure}

 After training, correlation matrices for the signal and background get stored that tell us the correlation among the input variables (shown in Figure~\ref{fig:cormatrix}). 

\begin{figure}[htb]
\begin{center}
\epsfig{file=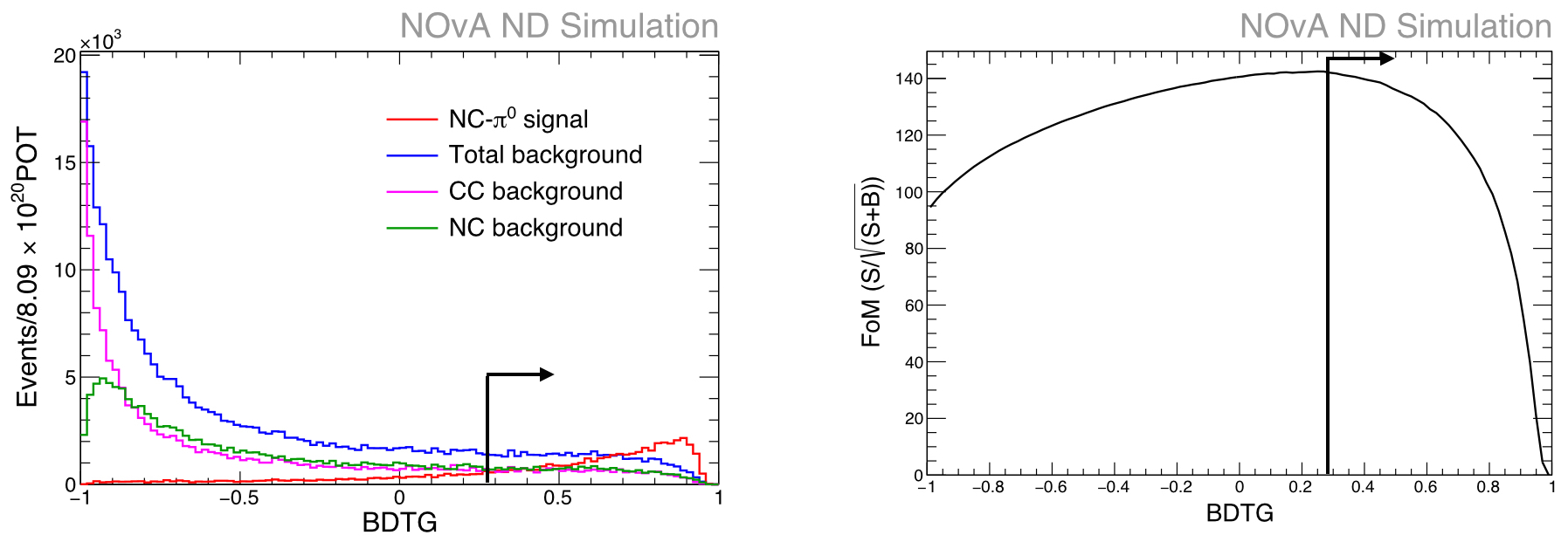,height=2.0in, width=0.8\textwidth}
\caption{(Left): BDTG distribution for the signal (Red) and background (Blue) events with CC and NC background.}
\label{fig:bdtg}
\end{center}
\end{figure}

When we train the BDTG, each variable gets a weight. Using that weight, we evaluate an output (BDTG output) (shown in Figure~\ref{fig:bdtg} (Left)). The BDTG output distribution for the signal and background is with the pre-selection (fiducial, containment, ReMId $<$ 0.36 and Reco K.E. $>$ 0.5 GeV). To select the signal events, we apply a cut on the BDTG based on its FOM (shown in Figure~\ref{fig:bdtg} (Right)). The FOM is maximized at 0.27.

\section{Results}

Table~\ref{tab:eventcount} shows the event counts for the signal and background at each cut level. The event numbers correspond to  $~8.09\cdot10^{20}$ POT. Using the BDTG, we reject almost 98\% of the background w.r.t the background with fiducial and containment. The total background, after all the cuts, is composed of 53\% NC events and 47\% CC events. The signal is mostly from Deep Inelastic scattering (DIS) which forms 60\% of the total signal events and a remaining 27\% is resonance and 11\% is coherent. The signal efficiency  is 48\% w.r.t the signal events with fiducial and containment cuts and 15\% w.r.t to signal events with fiducial cut only. 

\begin{table}[b]
\begin{center}
\begin{tabular}{l|cccccc}  
Cuts &  Signal (S) & Background (B)  &  
S/B & FOM  \\ \hline 

 Fiducial+Containment  &   91007     &     1389929      &     0.06  &  74.8   \\
 + ReMId $<$ 0.36 &  85829     &     876124      &  0.09 & 87.5 \\ 
  + Reco K.E. $>$ 0.5 GeV &  55192     &   285675      &  0.2 & 94.5 \\ 
  + BDTG $>$ 0.27  &  43634     &   44361      &  0.98 & 147.09 \\ \hline
\end{tabular}
\caption{Number of the signal and background events at each cut level.}
\label{tab:eventcount}
\end{center}
\end{table}

\begin{figure}[htb]
\begin{center}
\epsfig{file=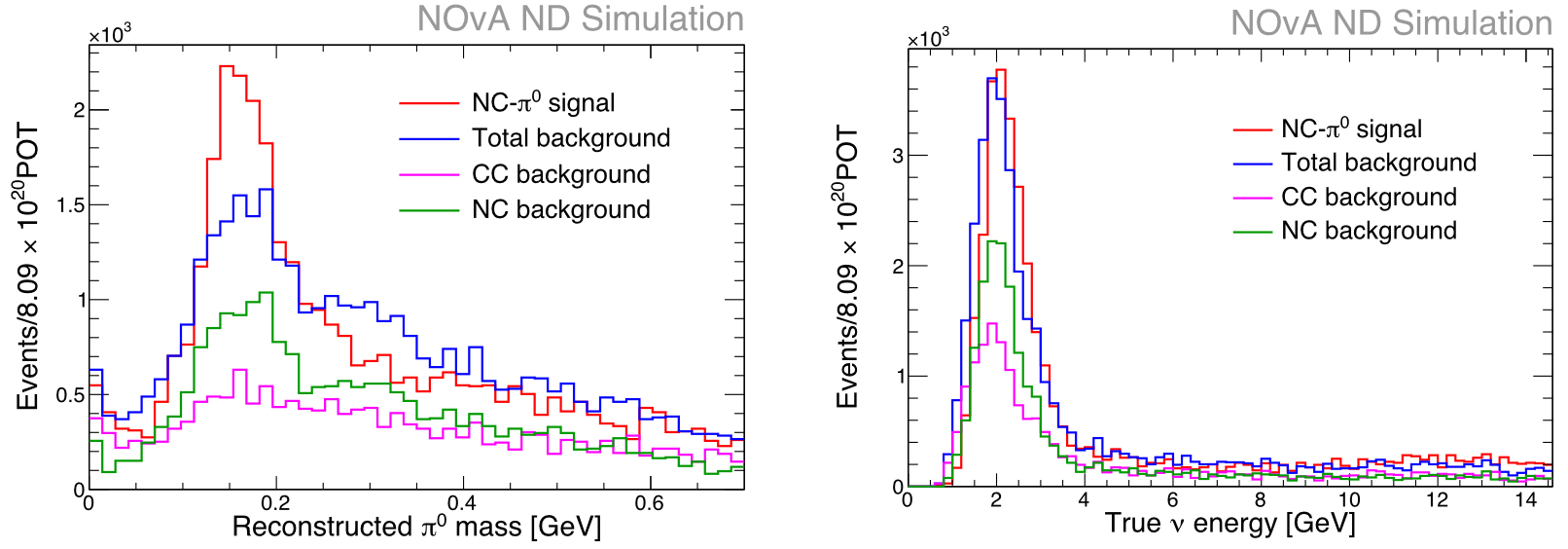,height=2.0in, width=0.8\textwidth}
\caption{(Left)Distribution of the reconstructed $\pi^{0}$ mass and (Right) Distribution of true $\nu$ energy for the signal (red) and background (blue) events.}
\label{fig:massandenergy}
\end{center}
\end{figure}

Figure~\ref{fig:massandenergy} shows the reconstructed $\pi^{0}$ mass and true $\nu$ energy distribution for the signal and background events with the pre-selection and BDTG $>$ 0.27. The reconstructed $\pi^{0}$ mass is calculated using prong1 and prong2 energy and cosine of angle between them. The dominant background component comes from the NC interactions (green) and gives a peak very similar to the signal peak. The NC background is mostly due to the low energy pions (K.E. $<$ 0.5 GeV) which are excluded in the signal. Study is going on to see if  some of the low energy pions can be included in the signal. 

\begin{figure}[htb]
\begin{center}
\epsfig{file=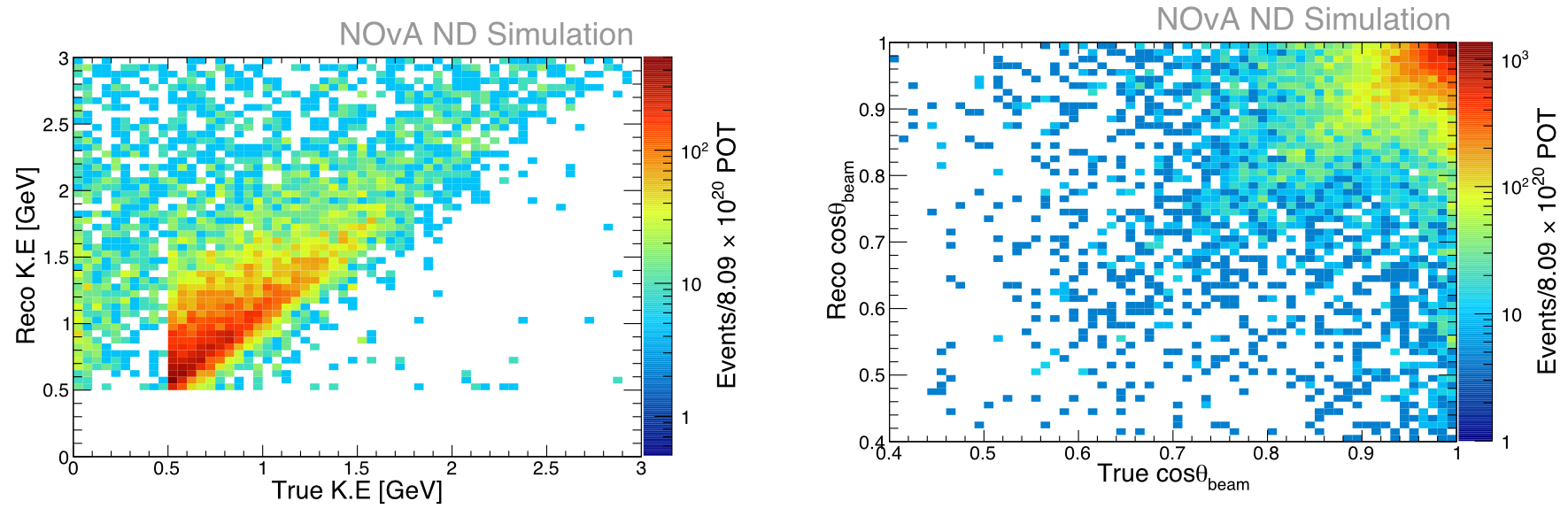,height=2.0in, width=0.8\textwidth}
\caption{(Left) Reconstructed vs true $\pi^{0}$ kinetic energy and (Right) Reconstructed vs true $\pi^{0}$ angle w.r.t beam.}
\label{fig:recovstrue}
\end{center}
\end{figure}

Figure~\ref{fig:recovstrue} shows the reconstructed vs true distribution for kinetic energy and $\pi^{0}$ angle w.r.t beam. Both the plots are drawn using logarithmic scale where we see a diagonal distribution of events. Note that in Reco vs true kinetic energy plot, there are no events below 0.5 GeV as we include a cut on the Reco K.E.  in the pre-selection.

\section{Cross Section and Final State Interaction (FSI) systematics}

We have seen from the event counts that the BDTG variable is powerful in rejecting the background but one can infer from the distributions in Figure~\ref{fig:massandenergy} that the signal and background events contribute almost equally even after all the selection cuts. We still want to reduce the background further to get a cleaner $\pi^{0}$ mass peak for instance. One idea is to choose a higher cut value on the BDTG distribution rather than the one which maximizes the FOM (BDTG $>$ 0.27). For this purpose, we look at the effect of systematic uncertainties on the BDTG distribution to optimize the event selection.

 \begin{figure}[H]
\begin{center}
\epsfig{file=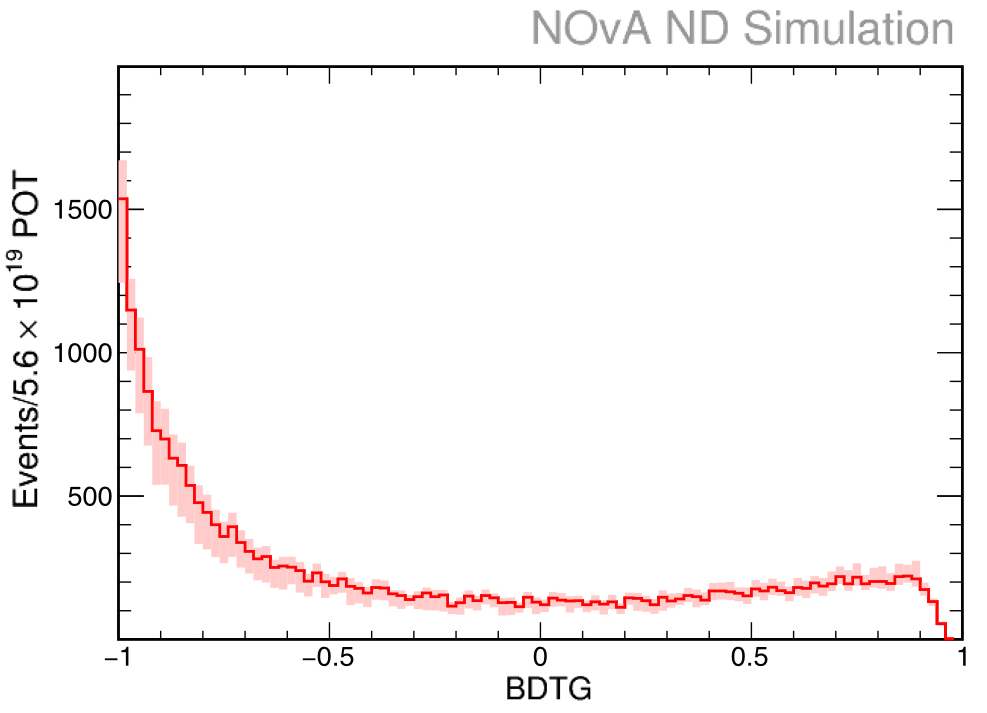,height=2.0in, width=0.5\textwidth}
\caption{BDTG distribution (Red) for the total number of events (signal and background) with $\pm$1 $\sigma$ error band .}
\label{fig:systematic}
\end{center}
\end{figure} 

We have looked at the effect of cross-section systematic uncertainties on the BDTG distribution (shown in Figure~\ref{fig:systematic}). The error band represents $\pm$1 $\sigma$ uncertainty. We use "Multi-universe" approach, where all the GENIE parameters can be varied at a same time, to treat the cross-section modeling systematic uncertainties. Figure~\ref{fig:systematic} shows that the error band stays almost constant in the BDTG range 0.2-0.7, indicating we are not biasing our selection too much by choosing a higher cut value for the BDTG. We are exploring other ways to optimize the signal selection.

\section{Summary}

This is a preliminary study where we used ND MC datasets with high statistics (x4 than the ND data). Multivariate algorithms are studied and we trained an algorithm (BDTG) to use it as a final event selector. We have developed an approach that is 48\% efficient at selecting the signal NC interactions with an energetic $\pi^{0}$ in the final state and rejects 98\% of the background interactions. The dominant background remainings are 53\% NC background and 47\% CC background events and we are investigating data-driven techniques to constrain these backgrounds. We are exploring to make the differential cross-section measurement with the NOvA ND.

\section{Acknowledgements}

NOvA is supported by the US Department of Energy; the US National Science Foundation; the Department of Science and Technology, India; the European Research Council; the MSMT CR, Czech Republic; the RAS, RMES, and RFBR, Russia; CNPq and FAPEG, Brazil; and the State and University of Minnesota.  We are grateful for the contributions of the staffs of the University of Minnesota module assembly facility and NOvA FD Laboratory, Argonne National Laboratory, and Fermilab. Fermilab is operated by Fermi Research Alliance, LLC under Contract No. DeAC02-07CH11359 with the US DOE.


\begin{thebibliography}{99}





\bibitem{Deborah}
Deborah A. Harris, The State of the Art of Neutrino Cross Section Measurements, Fermilab-Conf-15-254-ND.

\bibitem{Kevin}
Kevin S. McFarland, Neutrino Interactions, Conf:C06-08-08.

\bibitem{Formaggio}
J. A. Formaggio, From eV to EeV: Neutrino Cross Section Across Energy Scales, Fermilab-Pub-12-785-E.

\bibitem{Arevalo}
Alexis A. Aguilar-Arevalo et al., Measurement of $\nu_{\mu}$ and  $\overbar{\nu}_{\mu}$ induced neutral current single $\pi^{0}$ production cross-section on mineral oil at $E_{\nu}$ $\approx$ 1 GeV, Phys.Rev. D81 (2010) 013005. 

\bibitem{Adamson}
P. Adamson et al., First measurement of electron neutrino appearance in NOvA, Phys. Rev. Lett. 116 (2016) no.15, 151806

\bibitem{AdamsonI}
P. Adamson et al., First measurement of muon-neutrino disappearance in NOvA, Phys. Rev. D 93, 051104 (2016).

\bibitem{AdamsonII}
P. Adamson et al., Measurement of neutrino mixing angel $\theta_{23}$ in NOvA, Phys. Rev. Lett.  118 (2017) no.15, 151802.

\bibitem{Talaga}	
R. L. Talaga et al., Report No. FERMILAB-PUB-15-049- ND-PPD	

\bibitem{Fernandes}
L. A. F. Fernandes and M. M. Oliveira, Pattern Recognition 41, 299 (2008).
	
\bibitem{Andreopoulos}
C. Andreopoulos, C. Barry, S. Dytman, H. Gallagher, T. Golan, R. Hatcher, G. Perdue, J. Yarba., The GENIE Neutrino Monte Carlo Generator: Physics and User Manual, Fermilab-FN-1004-CD.

\bibitem{Hoecker}
A. Hoecker, P. Speckmayer, J. Stelzer, J. Therhaag, E. von Toerne, H. Voss, TMVA - Toolkit for Multivariate Data Analysis, arXiv:physics/0703039 [physics.data-an] [2009].

\bibitem{Aurisano}
A. Aurisano, A. Radovic, D. Rocco, A. Himmel, M.D. Messier, E. Niner, G. Pawloski, F. Psihas, A. Sousa, P. Vahle, JINST 11 (2016) no.09, P09001.

\end{thebibliography}
\end{document}